\documentclass{iau} 
\usepackage{graphicx}
\usepackage[T1]{fontenc}
\usepackage{natbib}

\title[NIR Spectroscopic Studies of $z\sim2$ Galaxies] 
{Near-IR Spectroscopic Studies of Galaxies at $z\sim 1-3$} 

\author[Irene Shivaei]   
{Irene Shivaei
}

\affiliation{Steward Observatory, University of Arizona, Tucson, AZ 85721, USA \\ Hubble Fellow \\email: {\tt ishivaei@email.arizona.edu} 
}

\pubyear{2019}
\volume{352}  
\jname{Uncovering Early Galaxy Evolution in the ALMA and JWST Era}
\editors{Elisabete da Cunha, Jacqueline Hodge, Jose Afonso, Laura Pentericci}
\begin{document}

\maketitle

\begin{abstract}
ISM comprises multiple components, including molecular, neutral, and ionized gas, and dust, which are related to each other mainly through star formation -- some are fuel for star formation (molecular gas) while some are the products of it (ionized gas, dust). To fully understand the physics of star formation and its evolution throughout cosmic time, it is crucial to measure and observe different ISM components of galaxies out to high redshifts. I will review the current status of near-IR studies of galaxies during the peak of star formation activity ($z\sim 1-3$). Using rest-frame optical emission lines, we measure dust, star formation, and gaseous properties of galaxies. {\em JWST} will advance such studies by probing lower luminosities and higher redshifts, owing to its significantly higher sensitivity. Incorporating ALMA observations of cold dust and molecular gas at $z>1$ will give us a nearly complete picture of the ISM in high-redshift galaxies over a large dynamic range in mass.
\keywords{galaxies: high-redshift, galaxies: abundances, galaxies: evolution, galaxies: general, galaxies: ISM, dust, extinction}
\end{abstract}

\firstsection 

\section{Introduction}
\begin{figure}[b]
\begin{center}
 \includegraphics[width=.8\textwidth]{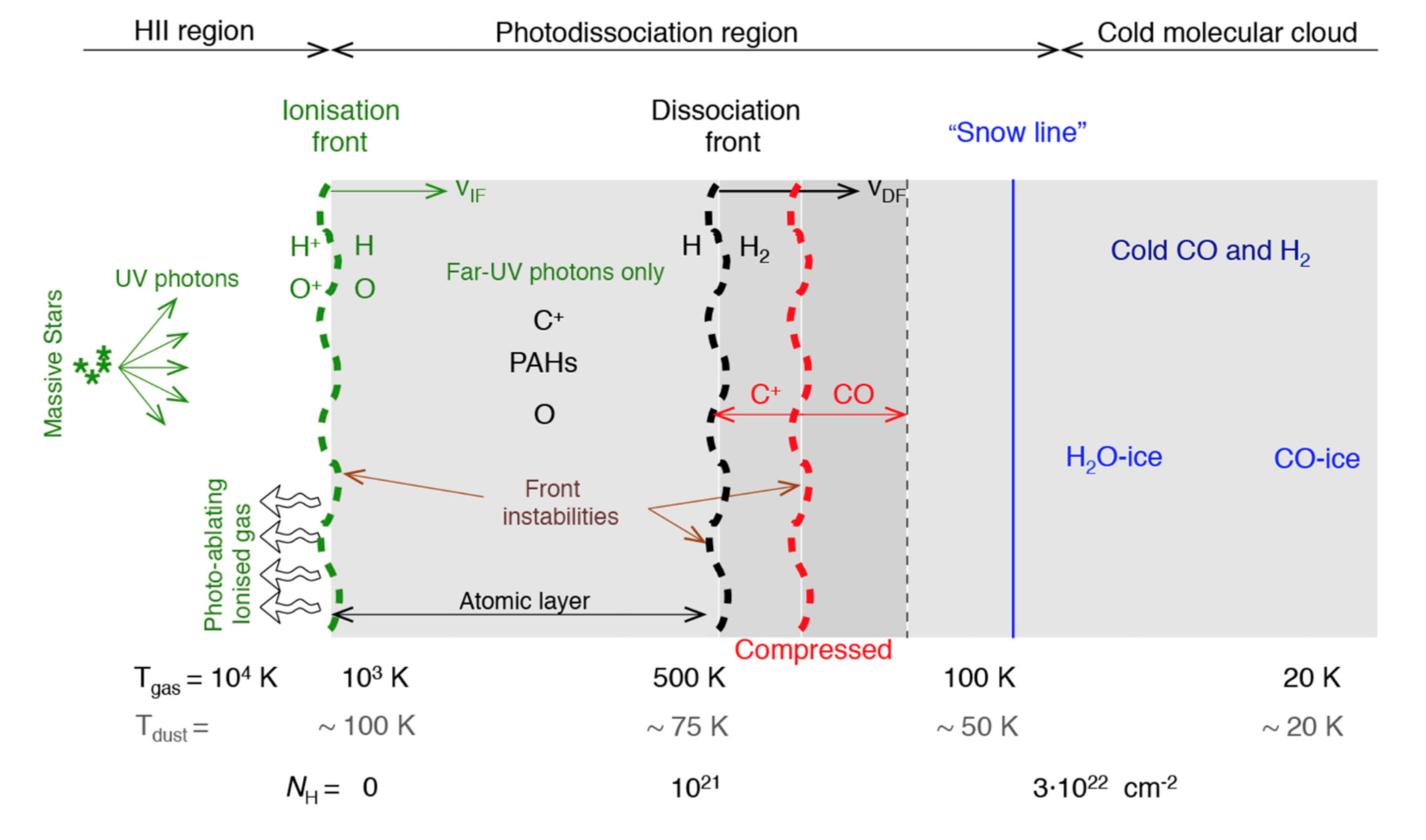} 
 \caption{Structure of a UV-irradiated molecular cloud edge \citep{goicoechea16}}
   \label{fig:ism}
\end{center}
\end{figure}

The interstellar medium (ISM) comprises various components, including molecular, neutral, and ionized gas, large dust grains in thermal equilibrium with their surrounding, and small dust grains, such as PAHs, heated by single photons (Figure~\ref{fig:ism}). The different components of ISM are related to each other mainly through star formation -- some are fuel for star formation (molecular gas) while some are the products of it (ionized gas, dust). Therefore, to fully understand the physics of star formation, its evolution, and its connection to metal and dust enrichment, it is crucial to trace various ISM components through a range wavelengths. The ionized phase of ISM, which is the focus of this review, is gas photoionized by the energetic photons of hot stars, and, if present, AGN. It has a temperature of ${\rm T}\sim 10^{4}$\,K and can be in dense or diffuse regions with densities of $n_{\rm H}\sim 0.3-10^4\,{\rm cm^{-3}}$ \citep{draine11}. Some the main diagnostics of the properties of ionized gas are the nebular emission lines in the rest-frame optical spectra of galaxies, including hydrogen Balmer lines, H$\alpha$, H$\beta$, and atomic fine-structure lines of [O{\sc ii}], [O{\sc iii}], [N{\sc ii}], and [S{\sc ii}]. 

In this review, I focus on the advances of our understanding in ISM physics at $z\sim 1-3$, corresponding to $\sim 2-6$ billion years after Big Bang. Redshifts of $z\sim 1-3$ are an important era to study star formation activity and metal and dust enrichment in the history of universe, as galaxies in that era were in the process of assembling most of their stellar mass and quasar activity was at its peak \citep{madau14}. At these redshifts, the optical nebular emission lines are redshifted to near-IR wavelengths.
Owing to the high near-IR terrestrial background, obtaining rest-frame optical spectra of large samples of galaxies at $z\sim 2$ has been very challenging prior to the advent of multi-object near-IR spectrographs on 10-m class telescopes.
As a result, the existing near-IR surveys were limited to small samples using the classical single-object long-slit instruments \citep[e.g.,][]{erb06a} or larger samples of low resolution grism data \citep[e.g.,][]{brammer12}.
The multiplexing capability and high sensitivity of revolutionary instruments such as MOSFIRE on the 10-m Keck (first light in 2012; \citealt{mclean12}), KMOS on the 8.2-m VLT (first light in 2012; \citealt{sharples13}), FMOS on the 8.2-m Subaru (first light in 2008; \citealt{kiruma10}), and LUCI on the 8.4-m LBT (first light in 2008; \citealt{ageorges10}), enabled the first statistically large studies of ionized ISM at high redshifts.
Among these large spectroscopic surveys of $z\sim 1-3$ galaxies are KBSS-MOSFIRE \citep{steidel16,strom17} and MOSDEF \citep{kriek15} with respectively $\sim 1100$ and $\sim 1500$ MOSFIRE spectra of galaxies at $z\sim 1.5-3.5$, KMOS-3D with KMOS near-IR integral field spectra of over 700 galaxies at $z\sim 0.6-2.7$ \citep{wisnioski15}, and FMOS-COSMOS with $\sim 1900$ FMOS spectra at $z\sim 1.6$ \citep{kashino19}.

In future, with {\em JWST} and its cutting-edge instruments, we will be able to build upon the current high-redshift near-IR surveys by probing fainter and higher redshift objects. The multi-object NIRSpec spectrograph on board of {\em JWST} will provide us with near-IR spectra out to 5\,$\mu$m, in the absence of terrestrial background and sky line contamination. NIRCam camera will obtain high resolution imaging out to 5$\mu$m, and with MIRI we will be able to acquire mid-IR photometry out to 28\,$\mu$m, with much higher sensitivity and spatial resolution compared to its predecessor, {\em Spitzer}.

\section{Optical emission line diagnostics}
In the rest-frame optical spectra, there are multiple strong nebular emission lines that can be used as diagnostics of the physical conditions of ionized gas (e.g., excitation and ionization properties, electron densities and temperatures, chemical abundances), stars (star-formation rate and production rate of ionizing photons), and dust (nebular reddening).
In this section, I will briefly review the observational analysis and diagnostic measurements of such properties using near-IR (rest-frame optical) spectra of $z\sim 2$ galaxies. In each section, the potential synergies with longer wavelength IR/submm facilities, such as {\em Sptizer}, {\em Herschel}, ALMA, and future {\em JWST} are also briefed. 

\subsection{Star formation rate and stellar ionizing radiation}
\label{sec:sfr}
The brightest emission line in the rest-frame optical spectra of galaxies is the H$\alpha$ Balmer line, which is emitted from the ionized gas around the most massive and hot stars, those with O and early-B spectral types. As the lifetime of these stars on the main sequence is very short ($\sim$10\,Myr), hydrogen nebular lines are considered as nearly instantaneous tracers of SFR. The conversion of H$\alpha$ luminosity to SFR has been studied by numerous authors and is widely used at low and high redshifts (see the review by \citealt{kennicutt12}). An important correction that needs to be applied to the observed H$\alpha$ luminosity to recover the total SFR, is dust attenuation correction. The reddening or color excess ($E(B-V)$) can be calculated using various methods, with the ratio of the two Balmer lines (i.e., Balmer decrement, see Section~\ref{sec:dust}) being the most direct one for nebular emission lines (Balmer decrement, see Section~\ref{sec:dust}).

\begin{figure*}
\begin{center}
 \includegraphics[width=.9\textwidth]{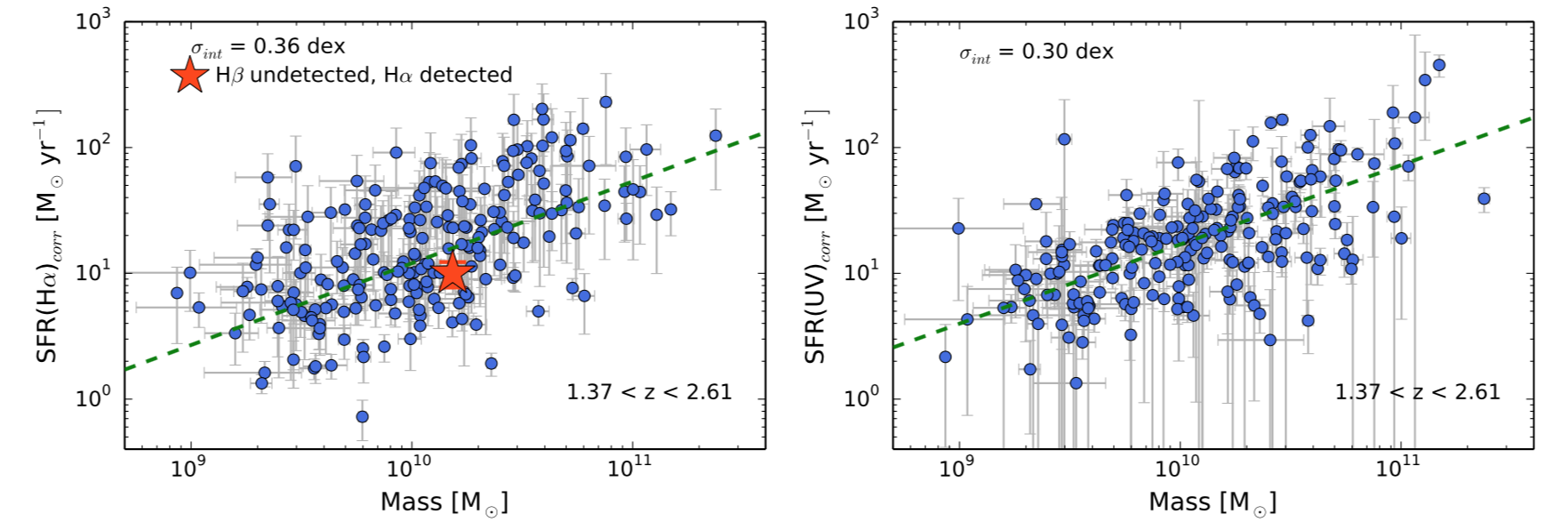} 
 \caption{SFR as a function of stellar mass for star-forming galaxies at $z\sim 2$. {\em Left}: SFR(H$\alpha$) -- corrected for dust attenuation assuming the Balmer decrement and a Cardelli et al. (1989) extinction curve -- vs. M$_*$. {\em Right}: SFR(UV) vs. M$_*$. The SFR is inferred from the UV luminosity at 1600\,\AA~and is dust corrected by the UV slope, both determined from the multiband photometry. The green lines show the regression lines fitted to the log(SFR)-log(M$_*$) relation. The measurement-subtracted scatter in SFR (i.e., the scatter after the subtraction of SFR and mass measurement uncertainties) is provided in the upper left corner of each plot. Figure from \citet{shivaei15b}.
 }
   \label{fig:shivaei15}
\end{center}
\end{figure*}

\begin{figure*}
\begin{center}
 \includegraphics[width=.75\textwidth]{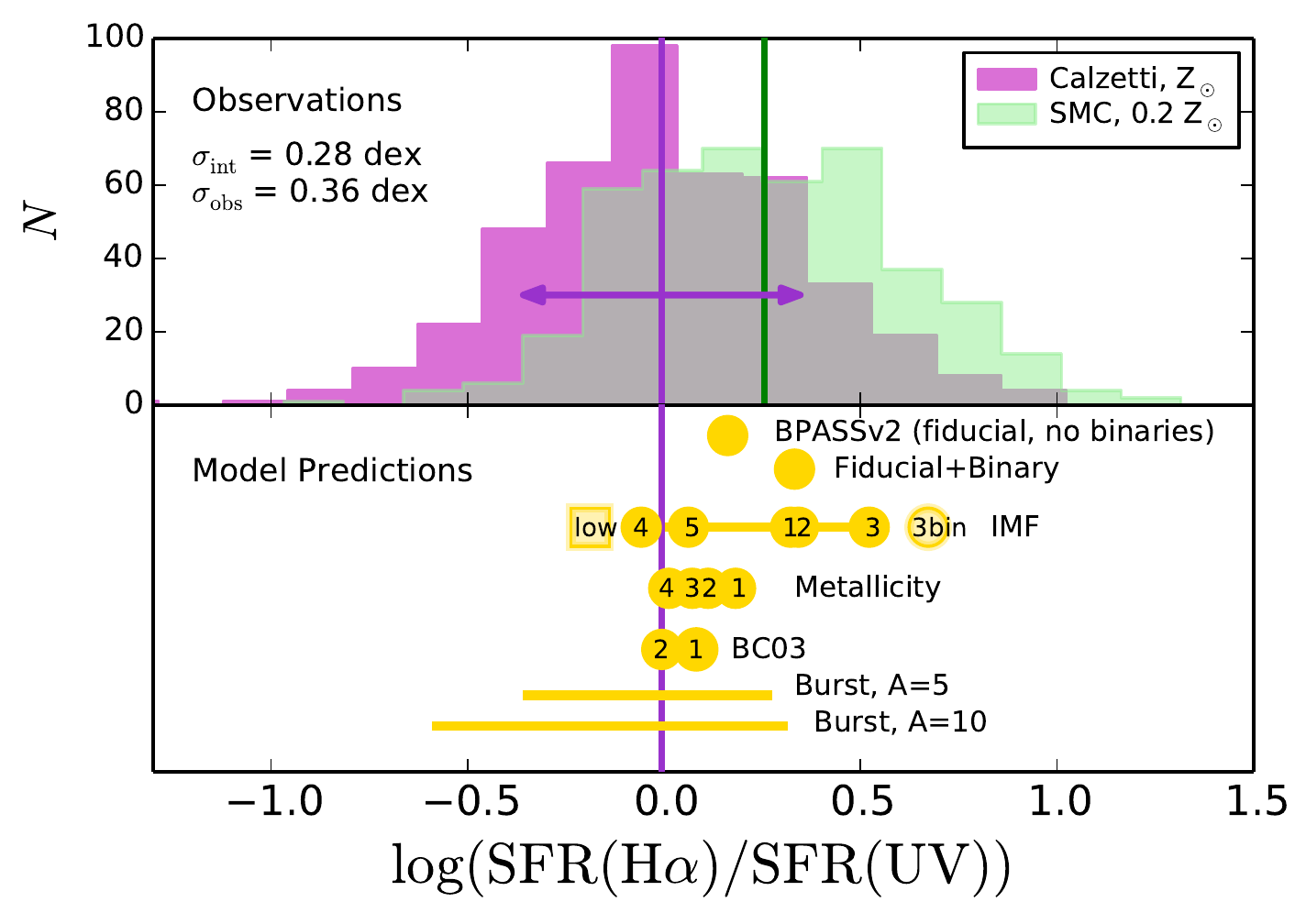} 
 \caption{{\em Top}: distribution of SFR(H$\alpha$)-to-SFR(UV) for a sample of $z=1.4-2.6$ galaxies, assuming the Calzetti attenuation curve for the UV dust correction in violet and an SMC curve in green. The observed and intrinsic (i.e., measurement-subtracted) scatters are 0.36 and 0.28\,dex for the Calzetti distribution and are shown with black and violet arrows, respectively. The vertical lines show the averages of the sample. Assuming an SMC curve systematically increases xion by $\sim$0.3\,dex, indicating the sensitivity of the SFR ratios to the assumed UV attenuation curve. {\em Bottom}: SFR(H$\alpha$)-to-SFR(UV) predictions from the stellar population models of BPASSv2 and BC03. Each row shows the variation of the SFRs ratio by changing the labeled quantity in the model, described in detail in Table 1 of \citet{shivaei18}. Figure is modified from \citet{shivaei18}.
}
   \label{fig:shivaei18}
\end{center}
\end{figure*}

Figure~\ref{fig:shivaei15} shows a comparison of the two SFR indicators, the attenuation-corrected H$\alpha$ and UV SFRs, in the SFR-M$_*$ plane at $z\sim2$. In principle, the two SFR indicators trace SFR on different timescales, and hence the difference in the scatter of the two SFR-M$_*$ relations (the $\sigma$ values on the corners of the plots in Figure~\ref{fig:shivaei15}) are thought to indicate recent star-formation burstiness. However, there are multiple factors in addition to a recent star formation burst that may cause deviations from the nominal value of the SFR(H$\alpha$)-to-SFR(UV) ratio in a galaxy. Figure~\ref{fig:shivaei18} shows the distribution of the ratio of dust attenuation corrected SFR(H$\alpha$)-to-SFR(UV) for a large sample of $z\sim 2$ galaxies from the MOSDEF survey \citep{shivaei18}. As shown, there is a large scatter in the observed values with an intrinsic scatter of 0.28\,dex. Part of the scatter and the uncertainty in such calculations are associated with uncertainties in the assumed dust attenuation curve (in this case, the Calzetti starburst curve versus the SMC extinction curve). Such uncertainties are discussed in more detail in Section~\ref{sec:dust}. Other than dust correction uncertainties, galaxy-to-galaxy variations in the IMF (the high-mass end slope or cutoff), stellar metallicity, and stellar rotation and binarity may also alter the rate of ionizing photons production at a given SFR. These effects are shown in the lower panel of Figure~\ref{fig:shivaei18}, calculated by varying different parameters in the stellar population synthesis models.
The details of the models can be found in \citet{shivaei18}.
The important take-away point is that although Balmer lines are tracers of the most recent SFR, their conversion to SFR should be taken with caution, particularly for high-redshift systems, which may have different stellar populations (e.g., lower stellar metallicities and more intense and/or harder ionizing radiation at a given SFR) compared to that of local galaxies.

\begin{figure*}
\begin{center}
 \includegraphics[width=.6\textwidth]{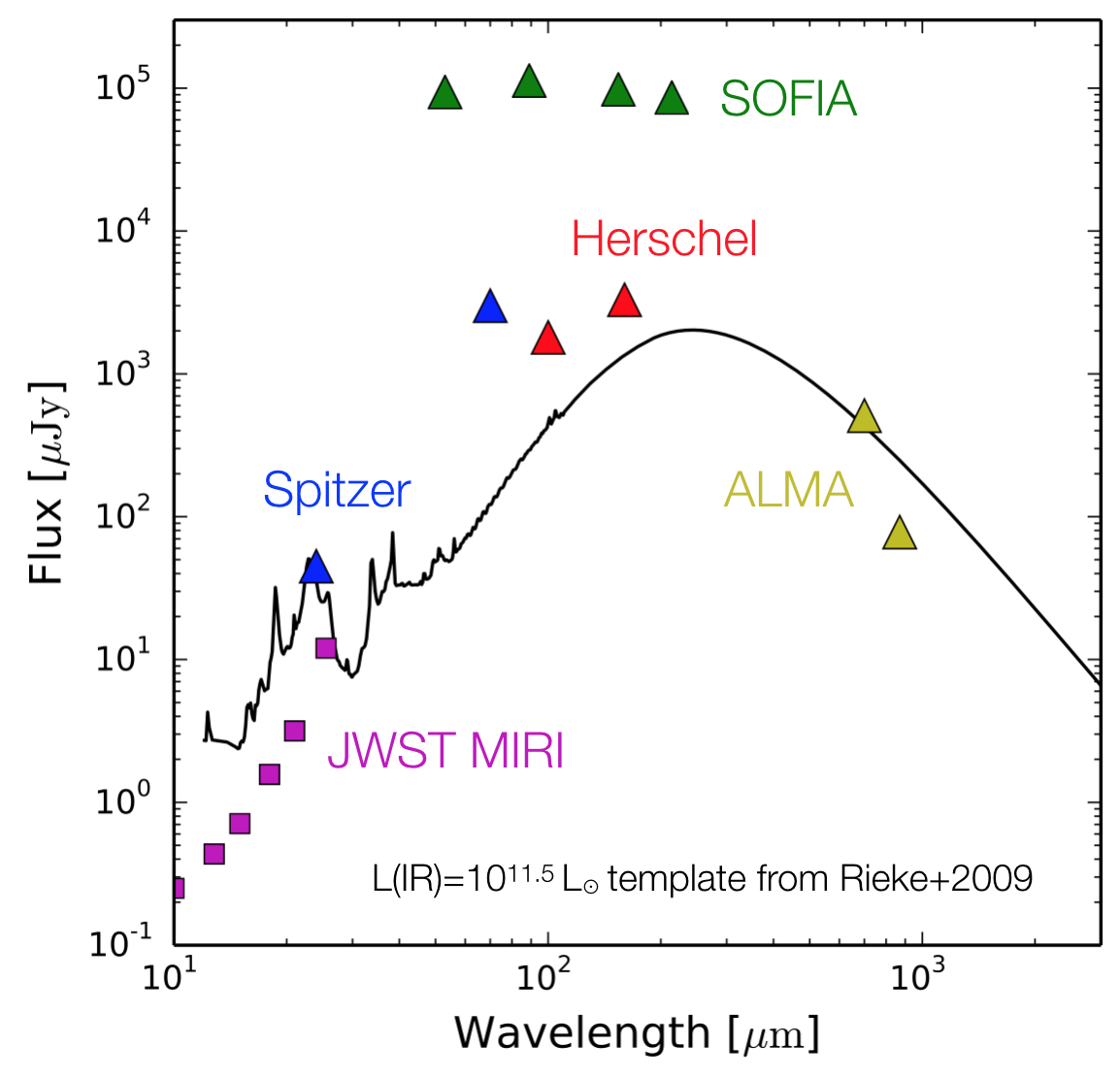} 
\caption{$3\sigma$ detection limits for 1-hour integrations with the major previous, current, and future IR and sub-mm facilities. The limits for MIPS 70 and 24$\mu$m are calculated based on the confusion limits from \citet{frayer06}. The black SED is the 10$^{11.5}$\,$L_{ \odot}$ template of \citet{rieke09}, represented as the IR emission of a  {\em typical} $z\sim 2$ galaxy (based on the IR luminosity function of \cite{magnelli11}, L(IR)$^*$ is $10^{11.83\pm 0.13}\,L_{\odot}$ at $z\sim 2$).
}
   \label{fig:irsed}
\end{center}
\end{figure*}

\begin{figure*}
\begin{center}
 \includegraphics[width=.75\textwidth]{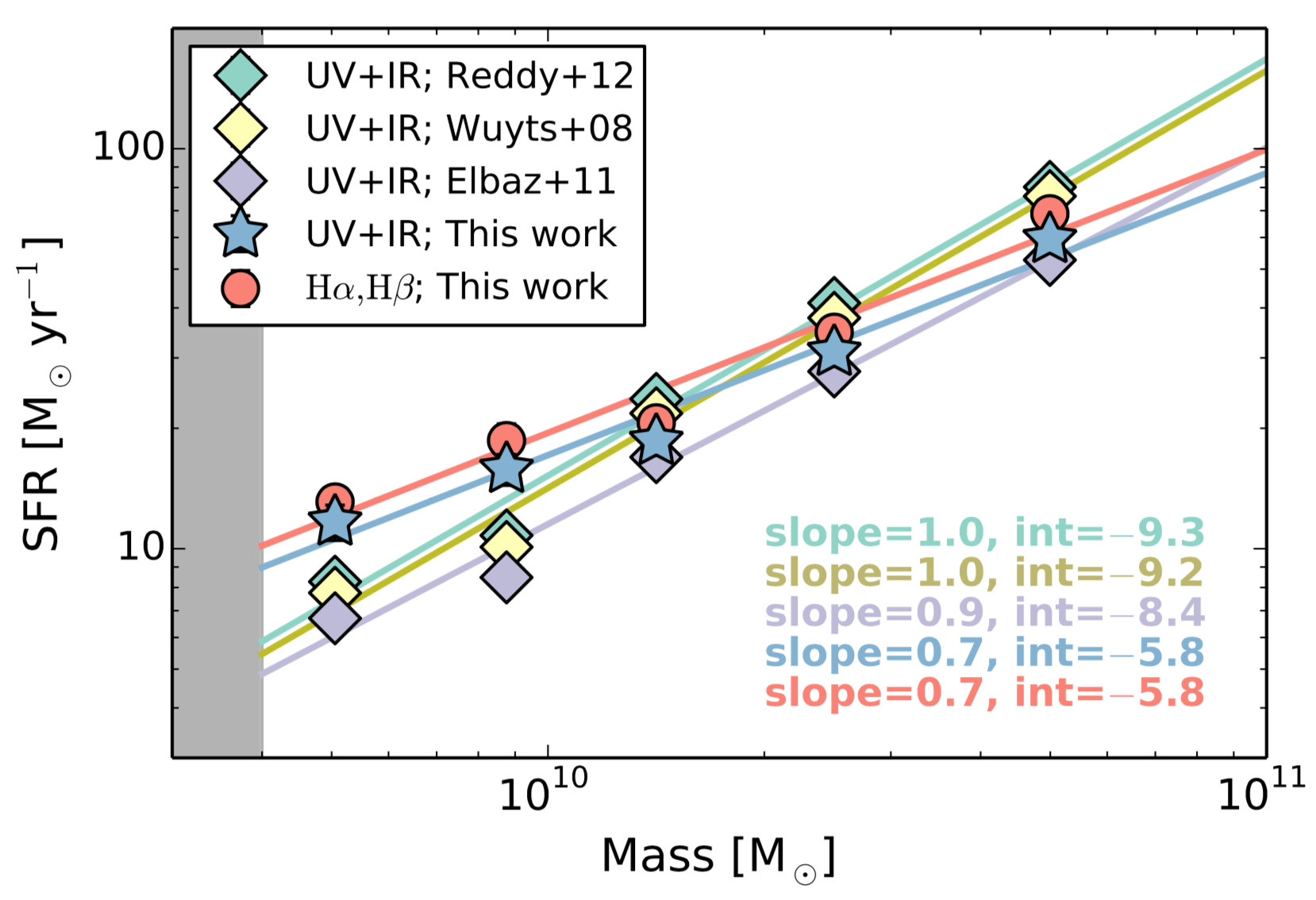} 
\caption{Star-formation rate as a function of stellar mass at $z\sim 2$. Orange circles are averages of dust-corrected H$\alpha$ SFR in each bin of stellar mass. Other symbols show the sum of SFR(UV) at 1600\AA~and SFR(IR) derived from 24$\mu$m stacks using different conversions. Blue stars adopt the mass-dependent conversion of \citep{shivaei17}. Cyan, yellow, and purple diamonds are from single-value conversions of \citet{reddy12a}, \citet{wuyts08}, and \citet{elbaz11}, respectively. The best-fit lines, slopes, and intercepts estimated from a simple linear least-squares regression to log(SFR) vs. log(M$_*$) are shown on the plot with respective colors. Figure from \citet{shivaei17}.
}
   \label{fig:shivaei17-msfr}
\end{center}
\end{figure*}

While dust-corrected H$\alpha$ SFRs are commonly used, the Balmer lines may miss optically thick star-forming regions \citep{shivaei16}. 
This is an important issue at redshifts of $z\sim 1-3$, where galaxies are more star forming (and hence, more dusty) at a given stellar mass compared to $z\sim 0$, and the obscured star formation dominates the total SFR at masses of $M_*\gtrsim 10^{9.5}$\,M$_{\rm \odot}$ \citep{whitaker17}. 
Therefore, to gain a full understanding of total star formation activity in galaxies at $z\sim 2$, both the obscured and unobscured components of star formation need to be measured.
{\em Spitzer} and {\em Herschel} opened a new window into measuring bolometric SFRs by allowing us to directly trace the obscured star forming regions \citep{reddy12a,whitaker14b,shivaei15a}. However, due to the low spatial resolution and sensitivity of PACS and SPIRE, {\em Herschel} studies at high redshifts are limited to dusty and IR-bright galaxies. Meanwhile, the higher sensitivity of {\em Spitzer}/MIPS 24$\mu$m detector enabled us to detect mid-IR emission in {\em individual typical} galaxies at $z\sim 2$ (Figure~\ref{fig:irsed}). MIPS 24$\mu$m traces the PAH emission at $z\sim 2$, which can be converted to total IR luminonisity (L(IR)) by using empirical conversions \citep{wuyts08,reddy12a} or IR templates \citep{ce01,rieke09,elbaz11}.  However, caution needs to be taken in using such conversions at low masses (${\rm M_*}<10^{10}$\,M$_{\rm \odot}$) as the PAH intensity scales with metallicity, such that at low metallicities (and low masses, owing to the mass-metallicity relation), the PAH-to-L(IR) ratio decreases by a factor of $\sim 2$ \citep{shivaei17}. Figure~\ref{fig:shivaei17-msfr} shows the good agreement between the SFRs inferred from dust-corrected H$\alpha$ luminosity with those derived from PAH emission using the mass-dependent calibration of \citet{shivaei17}. In the same figure, it is shown that using the single conversions of PAH-to-L(IR) from literature underestimates SFR at low masses, as such conversions are calibrated based on samples of massive and dusty galaxies. 

ALMA bands 6 and 7 trace the Rayleigh-Jeans (RJ) tail of the IR emission at $z\sim 1-3$, which can be used to constrain dust masses and possibly the total IR emission \citep{aravena16, dunlop17, franco18}. The disadvantages of these observations are a) due to the small field of view of ALMA, it is very time consuming (and not feasible) to obtain large surveys of IR emission of typical galaxies at high redshifts, and b) the conversion of the RJ emission to L(IR) is highly dependent on the assumed IR template, which is uncertain at high redshifts and low-metallicity regimes, as most of the existing templates are locally calibrated \citep{casey12, derossi18, schreiber18}.

The higher sensitivity and angular resolution of {\em JWST}/MIRI compared to {\em Spitzer}/MIPS 24$\mu$m\footnote{{\em JWST}/MIRI has $\sim 50$ times the sensitivity and 7 times the angular resolution of {\em Spitzer}/MIPS.} will enable us, for the first time, to obtain a nearly complete census of the PAH and IR emission at high redshifts, and to detect the 7.7$\mu$m feature in {\em individual} galaxies down to masses as low as $10^9$\,M$_{\rm \odot}$ (Figure~\ref{fig:irsed}). 

\subsection{Electron density and temperature}
\label{sec:ne}
The average electron density ($n_e$) may be estimated by comparing the line fluxes of two lines of the same ion, in which the lines have nearly the same excitation energies but different collision strengths and radiative transition probabilities. In this case, the relative population of the two levels, and hence the relative intensities of the two lines, depends on the density of electrons \citep{osterbrock06}. The two commonly-used electron density-sensitive doublets are [O{\sc ii}]$\lambda\lambda3726,2729$ and [S{\sc ii}]$\lambda\lambda6716,6731$, with their critical densities being comparable to the average ISM electron densities of galaxies. Studies at $z\sim 2$ have shown that electron densities inferred from [O{\sc ii}] and [S{\sc ii}] doublets, assuming an electron temperature of 10$^4$\,K, are an order of magnitude higher at $z\sim 2$ compared to the SDSS results at $z\sim 0$ \citep{sanders16a,strom17}.

\begin{figure*}
\begin{center}
 \includegraphics[width=.75\textwidth]{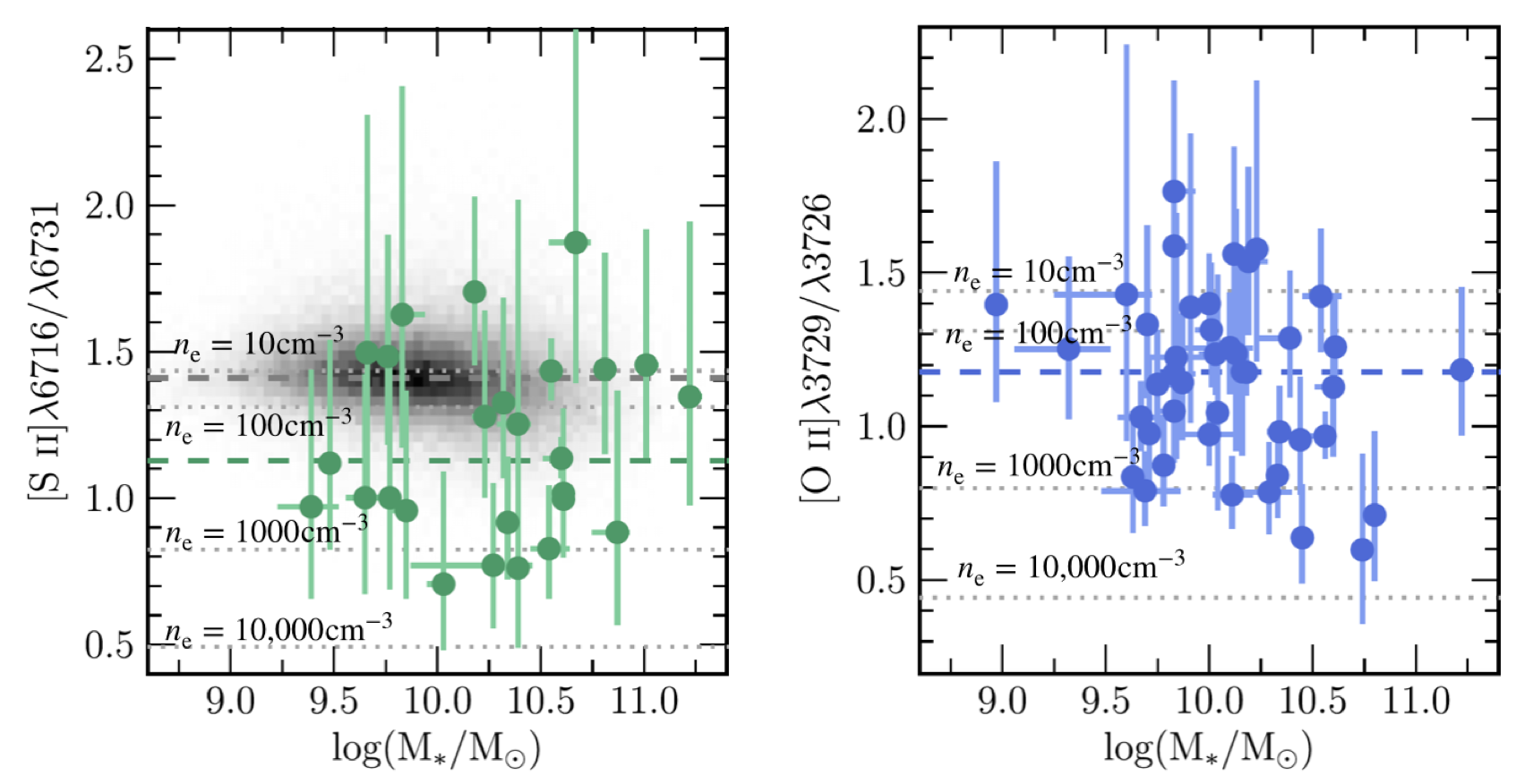} 
 \caption{The ratio of [S{\sc ii}]$\lambda\lambda6716,6731$ doublet (left) and [O{\sc ii}]$\lambda\lambda3726,2729$ doublet (right) as a function of stellar mass at $z\sim 2.3$ (green and blue points) and $z\sim 0$ (gray histogram). The green, blue, and gray lines show the median line ratios for the corresponding samples. Dotted horizontal lines show the line ratios corresponding to electron densities of 10, 100, 1000, and 10,000 cm$^-3$. Figure modified from \citet{sanders16a}.
}
   \label{fig:sanders16}
\end{center}
\end{figure*}

\label{sec:te}
Electron temperature ($T_e$) of the ionized gas may be estimated by using the ratio of the intensities of two lines of the same ion but with considerably different excitation energies, which makes the ratio strongly dependent on $T_e$. An example of such line ratios in the optical spectra are [O{\sc iii}] $\lambda4363$ auroral line to $\lambda4959$ and $\lambda5007$, and [N{\sc ii}]$\lambda5755$ to $\lambda6583$ and $\lambda6548$ \citep{osterbrock06}. The auroral lines are extremely week (e.g., [O{\sc ii}]$\lambda4363$ is typically 100 times weaker than [O{\sc iii}] $\lambda5007$ at low metallicities, and even weaker in solar and higher metallicities). The faintness of these lines makes them extremely hard to detect at high redshifts, resulting in fewer than 20 galaxies at $z>1$ that have $T_e$ measurements \citep{sanders19}. 

\subsection{Metallicity}
\label{sec:metal}

The most direct way of getting the abundances of ions is by taking the relative strength of their recombination lines to that of their Balmer lines. The strength of metal recombination lines are mildly dependent on electron temperature but due to the low elemental abundances, the metal recombination lines are very weak compared to hydrogen recombination lines, making this method practically limited to very bright HII regions.

The widely adopted method to determine chemical abundances in galaxies is using the relative ratio of the collisionally excited lines of elements to hydrogen recombination lines. The flux of collisionally excited lines depends on electron temperature and density -- once those are determined, one can infer the ionic abundances, and by assuming an ionization correction for the unobserved ions, the elemental abundances (metallicity) can be derived. In the optical spectra, the most commonly used auroral line to derive $T_e$ is [O{\sc iii}]$\lambda4363$, and the corresponding lines to measure the oxygen abundance ($12+\log{{\rm(\frac{O}{H})}}$) are [O{\sc iii}]$\lambda\lambda4959,5007$ and Balmer lines. 
However, as mentioned in Section~\ref{sec:te}, auroral lines that are used to estimate $T_e$ are intrinsically faint, making the direct $T_e$-based methods hard to conduct, particularly at high redshifts. As a result, studies rely on calibrations of strong emission line ratios to derive metallicities. Such calibrations are made empirically based on $z\sim 0$ galaxies or by using photoionization models.
The most commonly used strong line diagnostics are R23 ($\frac{\rm [O{\sc II}]\lambda3727+[O{\sc III}]\lambda\lambda4959,5007}{\rm H\beta}$), O3N2 ($\frac{\rm [O{\sc III}]\lambda\lambda4959,5007/H\beta}{\rm [N{\sc II}]\lambda6584/H\alpha}$), and N2 ([N{\sc ii}]$\lambda6584$/H$\alpha$; e.g., \citealt{kewley02,pp04}). The advantage of using the two latter diagnostics is the insensitivity to dust attenuation, as the dust attenuation is wavelength dependent and the constituent lines are close in wavelength.

\begin{figure*}
\begin{center}
 \includegraphics[width=.5\textwidth]{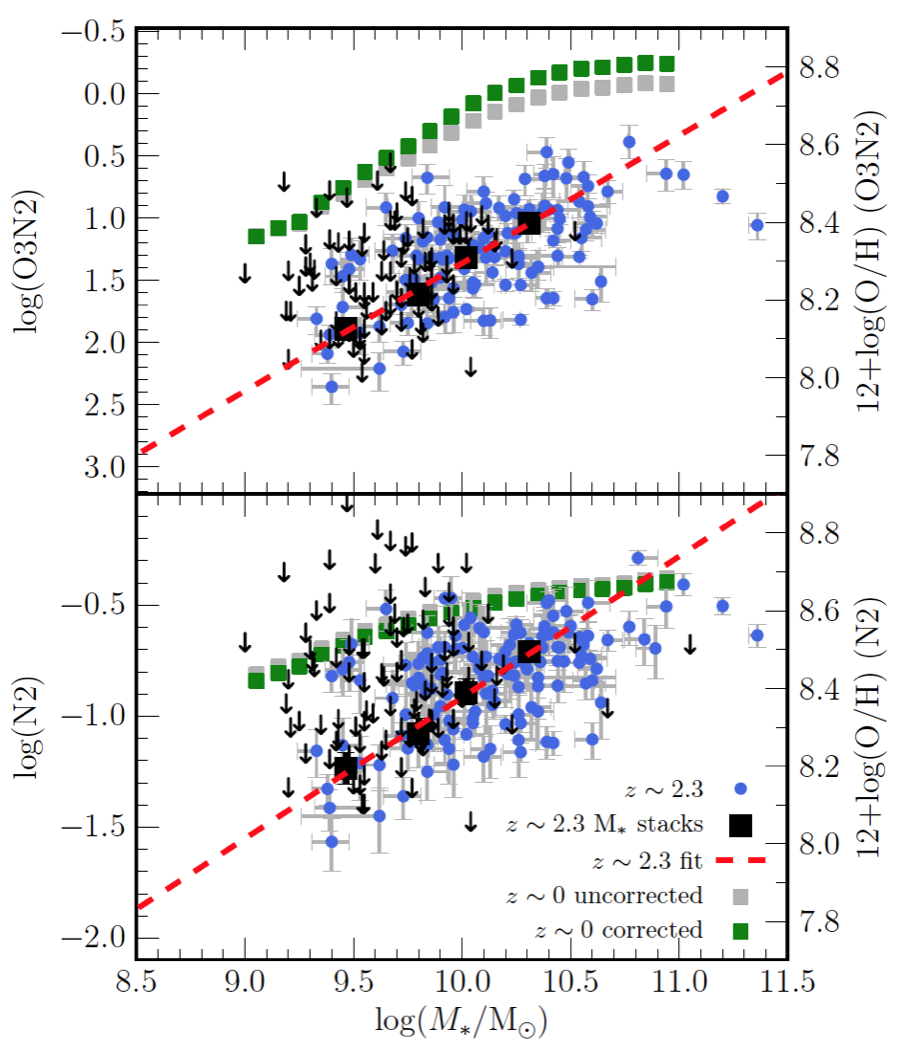} 
 \caption{ The emission line ratios $\frac{\rm [O{\sc III}]/H\beta}{\rm [N{\sc II}]/H\alpha}$ (O3N2; top) and [N{\sc ii}]/H$\alpha$ (N2; bottom) as a function of stellar mass at $z\sim 2.3$ and $z\sim 0$. The right y-axis shows the translation of the emission line ratios into metallicity. The black arrows display 3$\sigma$ limits in cases where one or more of the required lines is not detected. Black squares show stacks of both detections and non-detections in bins of stellar mass. The red dashed line is the best-fit linear relation to the stacks in each panel.
Figure from \citet{sanders18}.
}
   \label{fig:mz}
\end{center}
\end{figure*}

Studies show that galaxies at $z\sim 2$ have $\sim 0.3-0.5$\,dex lower metallicities at a given stellar mass compared to that at $z\sim 0$, depending on the metallicity indicator used (\citealt{strom17, sanders18}; the comparison of the $\frac{\rm [O{\sc III}]/H\beta}{\rm [N{\sc II}]/H\alpha}$, and [N{\sc ii}]/H$\alpha$ indicators between $z\sim 0$ and 2 are shown in Figure~\ref{fig:mz}).
However, a major concern is whether locally-calibrated strong line diagnostics still hold at high redshifts. For example, it is known that $z\sim 2$ galaxies have harder ionizing radiation \citep{steidel16,strom17}, higher ionization parameters \citep{kewley15,kashino17}, and higher electron density and ISM pressure \citep{liu08,bian10} at a fixed oxygen abundance, compared to local galaxies. Additionally, using the oxygen abundances may not be representative of the real metallicity of the gas, as the relative abundances of the chemical elements may vary with respect to the solar value at high redshifts. Indeed, studies show that $z\sim 1-3$ galaxies have lower Fe/H and higher O/H ratios compared to $z\sim 0$ galaxies \citep{steidel16,strom18,kriek19}.

These uncertainties and complications indicate the need for direct $T_e$ method calibrations at high redshifts, as well as independent methods of inferring metallicity to overcome the $T_e$ dependency. 
The far-IR ground-state fine structure lines provide independent diagnostics that are insensitive to electron temperature and dust obscuration \citep{nagao11,smith19}. ALMA observations of such lines at high redshifts can be joined with optical line diagnostics to study metal enrichment and directly calibrate metallicity diagnostics at high redshifts.

\begin{figure*}
\begin{center}
 \includegraphics[width=.5\textwidth]{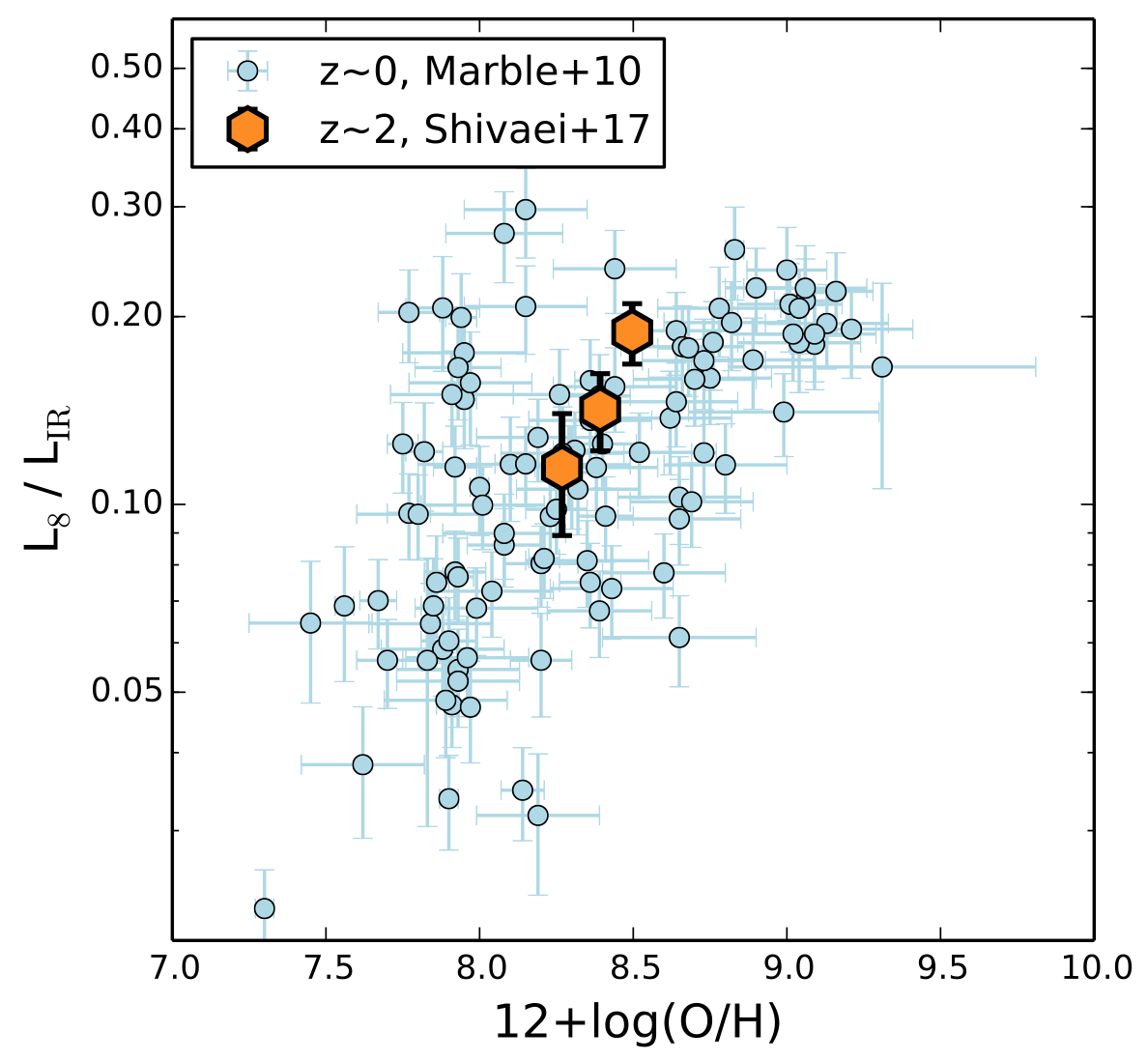} 
 \caption{ Relative strength of 7.7\,$\mu$m luminosity to total IR luminosity as a function of metallicity. Orange symbols are stacks of $z\sim 2$ galaxies in bins of metallicity, where the metallicity is derived from $\frac{\rm [O{\sc III}]/H\beta}{\rm [N{\sc II}]/H\alpha}$ diagnostic \citep{shivaei17}. The blue symbols are individual measurements at $z\sim 0$ \citep{marble10}. 
}
   \label{fig:shivaei17}
\end{center}
\end{figure*}

As a complementary tracer of metal enrichment, dust features such as the mid-IR PAH emissions may be valuable. Local and high-redshift studies have shown that the PAH intensity (defined as the strength of PAH 7.7$\mu$m feature to IR luminosity or SFR) drops below a certain oxygen abundance ($12+\log{{\rm(\frac{O}{H})}}\sim 8.1-8.2$ at $z\sim 0$ and $\sim 8.4-8.5$ at $z\sim 2$; Figure~\ref{fig:shivaei17}; \citealt{draine07b,engelbracht05,shivaei17}). The nature of this drop may be due to lower carbon abundances in low-metallicity environment, or harder ionizing radiation that destroys the PAH molecules. The future {\em JWST}/MIRI will enable us to trace the PAH features in large samples of $z\sim 1-2$ galaxies, down to at least an order of magnitude lower masses than have been studied previously at these redshifts \citep{elbaz11,shipley16,shivaei17}.

\subsection{Nebular Dust}
\label{sec:dust}
\begin{figure*}
\begin{center}
 \includegraphics[width=.9\textwidth]{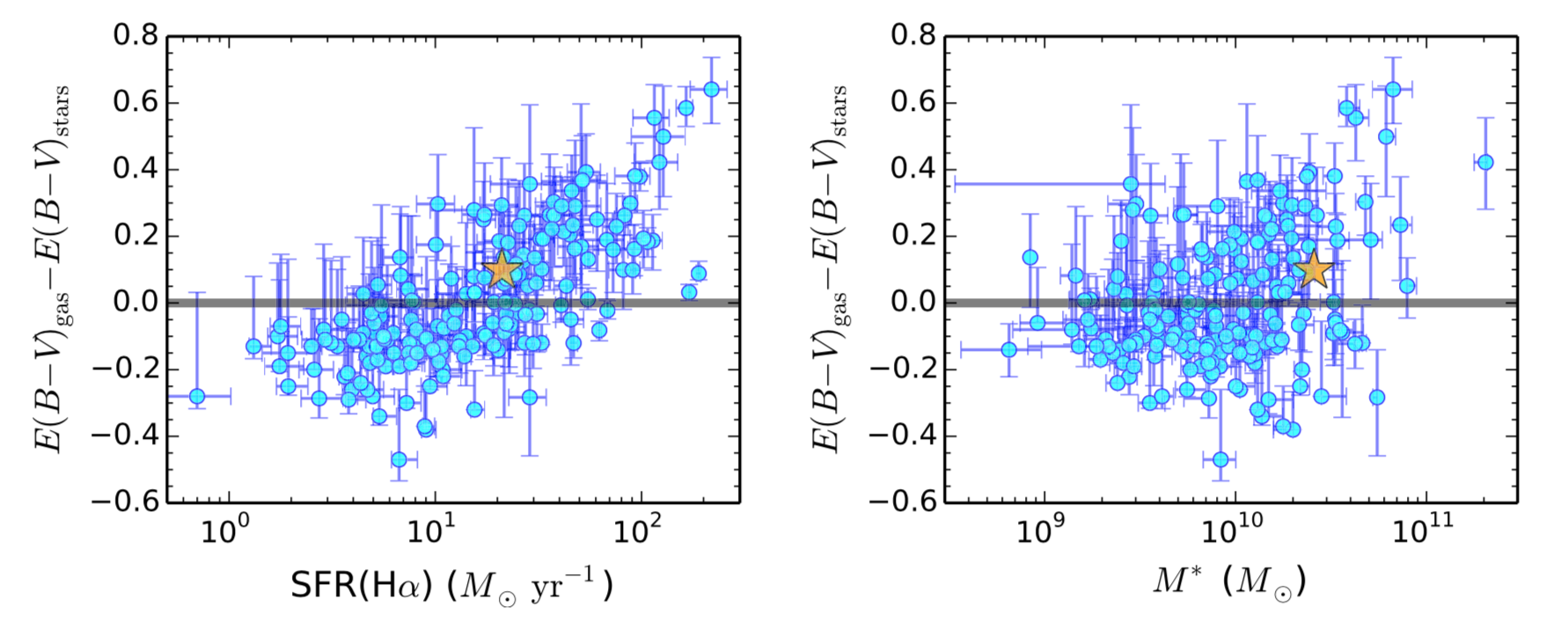} 
 \caption{Difference between gas and stellar continuum reddenings as a function of SFR (left) and stellar mass (right) for a sample of $\sim 200$ galaxies at $z\sim 2$. The solid lines indicate no difference between the color excess of the nebular regions and the stellar continuum. The yellow stars indicate the values for H$\beta$-undetected galaxies. Figure from \citep{reddy15}.
 }
   \label{fig:reddy15}
\end{center}
\end{figure*}

Balmer line ratios (e.g., Balmer decrement, H$\alpha$/H$\beta$) are almost insensitive to electron temperature and density, and hence, any deviation of the measured ratio from the theoretical intrinsic line ratio may be attributed to dust attenuation. Balmer optical depth is defined as $\tau_{\rm b}\equiv \ln({\frac{\rm H\alpha/H\beta}{2.86}})$, where 2.86 is the lines ratio for a Case B recombination with $T_e=10^4$\,K and $n_e=100\,{\rm cm^{-3}}$ \citep{osterbrock06}. Assuming a shape for the dust attenuation/extinction curve, Balmer optical depth indicates the nebular reddening ($E(B-V)_{\rm nebular}$), which is the reddening along the line of sight towards the ionized gas. Nebular reddening should be used to correct dust attenuation in nebular emission lines. On the other hand, $E(B-V)_{\rm stellar}$, inferred from the continuum indicators such as the UV slope, is advised to be used to correct the stellar continuum emission. Many studies in the literature have explored the relation between the two reddenings \citep{calzetti00,price14,shivaei15a,reddy15}. Owing to the stellar birth cloud dissipation and/or stars migrating from their parent molecular clouds, the differences between the mean optical depth probed by continuum and line photons may vary with SFR or specific SFR \citep{reddy15,theios19}. Therefore, the general consensus is that the relation between the two reddenings changes with SFR or specific SFR (Figure~\ref{fig:reddy15}).

At low redshifts, dust {\em extinction} curves have been determined along multiple sight lines of the Milky Way, LMC, and SMC \citep[e.g.,]{savage79,prevot84}. An extinction curve is determined for a simple system of a star with a foreground screen of dust. The dust screen absorbs and scatters photons out of the line-of-sight. Therefore, variations in extinction curves reflect differences in dust grain physics (the composition and size distribution).
On the other hand, {\em attenuation} curves are inferred for more complex systems, such as galaxies, in which the geometry of dust with respect to the emitting source also plays a role in determining the shape of the curve. In high redshift studies, the \citet{calzetti00} attenuation curve is often adopted, which is calculated based on a sample of local starburst galaxies.

\begin{figure*}
\begin{center}
 \includegraphics[width=.6\textwidth]{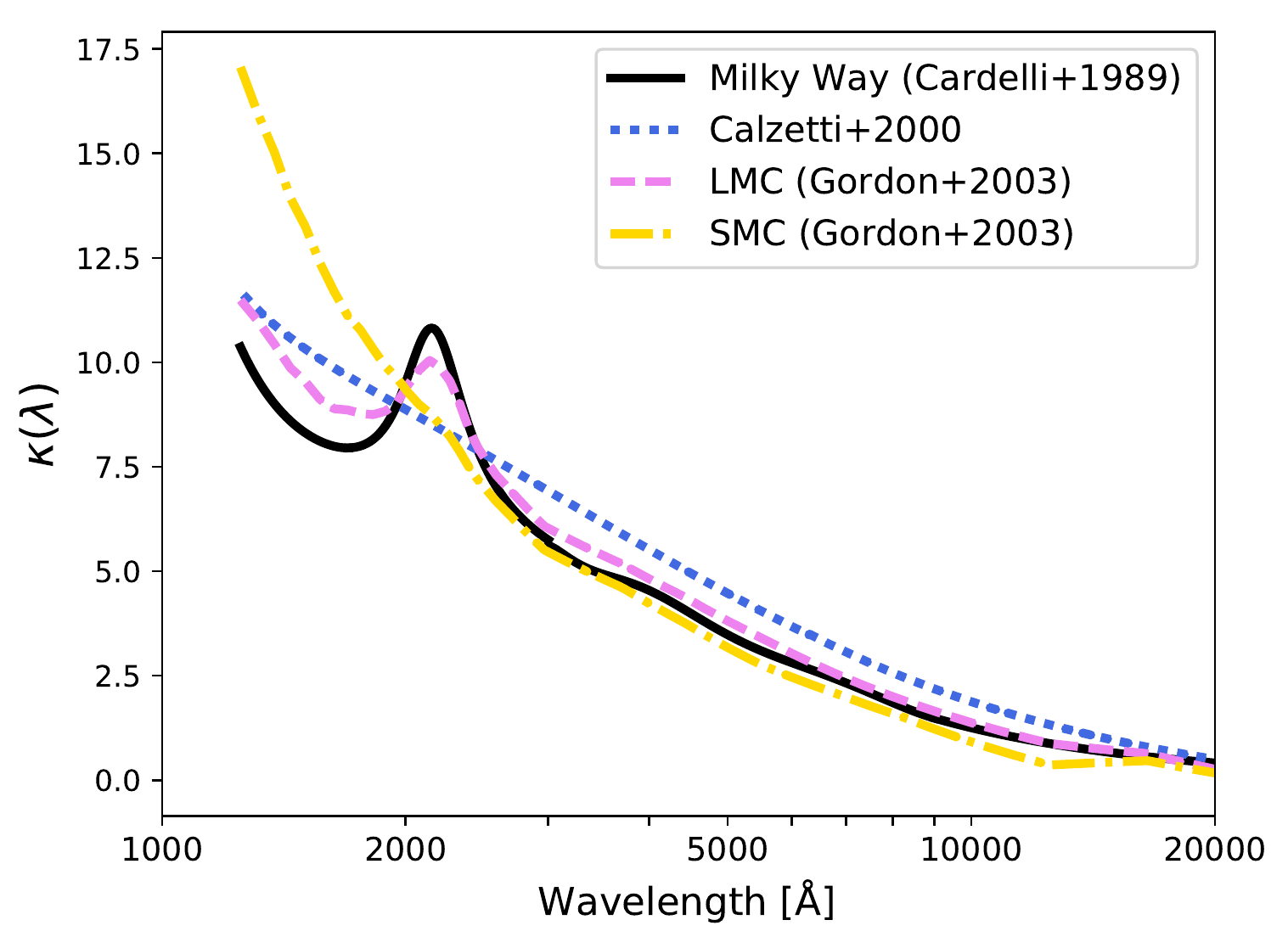} 
 \caption{Comparison of the commonly used dust extinction/attenuation curves. As evident, two of the most distinct differences between the curves are the steepness and the strength of a broad absorption feature at 2175\AA~(the UV bump).
 }
   \label{fig:attcurve}
\end{center}
\end{figure*}

The extinction and attenuation curves have profound differences. The two most distinct differences are the strength of the UV extinction bump at 2175\,\AA~and the UV slope of the attenuation curve. As shown in Figure~\ref{fig:attcurve}, the Milky Way and LMC extinction curves both have strong UV bumps, while the bump is absent in the SMC and the Calzetti curves. Also, the SMC curve has a much steeper slope compared to the other ones.
The underlying cause of these differences are not well known; it is speculated that as the SMC has the lowest metallicity of the three galaxies \citep[$12+\log({\rm O/H})\sim 8, Z\sim 0.2 Z_{\rm \odot}$][]{kurt98}, therefore it has a different grain size and composition distribution compared to the other two. These observations emphasize the importance of the galaxy-to-galaxy variations of the attenuation curves at high redshifts. 

Using Balmer decrement as an independent tracer of dust reddening, one can directly determine a dust attenuation curve by comparing the average SED of the ``more'' to ``less'' dusty galaxies with similar intrinsic (i.e., dust free) SED shapes (e.g., similar star formation histories, ages, etc.). This method requires large rest-optical spectroscopic samples \citep{calzetti94,battisti16}, and hence a challenging task at high redshifts \citep{reddy15}. Additionally, the normalization of the attenuation curve can be independently determined by incorporating IR data from {\em Spitzer} and {\em Herschel} and using an energy balance argument between the UV and IR emission \citep{calzetti00}.
Given the observed variations of the dust emission properties with gas-phase metallicity (e.g., the PAH intensity correlation with metallicity in Section~\ref{sec:metal}), it is also expected to see changes in the slope of the attenuation curve and the strength of the UV bump as a function of galaxy properties. Such studies require large enough spectroscopic datasets that can be divided into bins of metallicity, specific SFR, and stellar mass, to compare their inferred attenuation curves with each other.
In the future, with {\em JWST}, the dust attenuation properties of obscured and optically-thick systems can be independently studied by simultaneously observing multiple hydrogen recombination lines, including Paschen and Balmer series, and directly derive the nebular attenuation curve.

\bibliographystyle{apj}

\begin{thebibliography}{}
\expandafter\ifx\csname natexlab\endcsname\relax\def\natexlab#1{#1}\fi

\bibitem[{{Ageorges} {et~al.}(2010){Ageorges}, {Seifert}, {J{\"u}tte},
  {Knierim}, {Lehmitz}, {Germeroth}, {Buschkamp}, {Polsterer}, {Pasquali},
  {Naranjo}, {Gemperlein}, {Hill}, {Feiz}, {Hofmann}, {Laun}, {Lederer},
  {Lenzen}, {Mall}, {Mand el}, {M{\"u}ller}, {Quirrenbach}, {Sch{\"a}ffner},
  {Storz}, \& {Weiser}}]{ageorges10}
{Ageorges}, N., {Seifert}, W., {J{\"u}tte}, M., {et~al.} 2010, in Society of
  Photo-Optical Instrumentation Engineers (SPIE) Conference Series, Vol. 7735,
  Ground-based and Airborne Instrumentation for Astronomy III, 77351L

\bibitem[{{Aravena} {et~al.}(2016){Aravena}, {Decarli}, {Walter}, {Bouwens},
  {Oesch}, {Carilli}, {Bauer}, {Da Cunha}, {Daddi}, {G{\'o}nzalez-L{\'o}pez},
  {Ivison}, {Riechers}, {Smail}, {Swinbank}, {Weiss}, {Anguita}, {Bacon},
  {Bell}, {Bertoldi}, {Cortes}, {Cox}, {Hodge}, {Ibar}, {Inami}, {Infante},
  {Karim}, {Magnelli}, {Ota}, {Popping}, {van der Werf}, {Wagg}, \&
  {Fudamoto}}]{aravena16}
{Aravena}, M., {Decarli}, R., {Walter}, F., {et~al.} 2016, ApJ, 833, 71

\bibitem[{{Battisti} {et~al.}(2016){Battisti}, {Calzetti}, \&
  {Chary}}]{battisti16}
{Battisti}, A.~J., {Calzetti}, D., \& {Chary}, R.-R. 2016, ApJ, 818, 13

\bibitem[{{Bian} {et~al.}(2010){Bian}, {Fan}, {Bechtold}, {McGreer}, {Just},
  {Sand}, {Green}, {Thompson}, {Peng}, {Seifert}, {Ageorges}, {Juette},
  {Knierim}, \& {Buschkamp}}]{bian10}
{Bian}, F., {Fan}, X., {Bechtold}, J., {et~al.} 2010, ApJ, 725, 1877

\bibitem[{Brammer {et~al.}(2012)Brammer, van Dokkum, Franx, Fumagalli, Patel,
  Rix, Skelton, Kriek, Nelson, Schmidt, Bezanson, da~Cunha, Erb, Fan,
  F\'{o}rster~Schreiber, Illingworth, Labb\'{e}, Leja, Lundgren, Magee,
  Marchesini, McCarthy, Momcheva, Muzzin, Quadri, Steidel, Tal, Wake, Whitaker,
  \& Williams}]{brammer12}
Brammer, G.~B., van Dokkum, P.~G., Franx, M., {et~al.} 2012, ApJS, 200, 13

\bibitem[{Calzetti {et~al.}(2000)Calzetti, Armus, Bohlin, Kinney, Koornneef, \&
  Storchi-Bergmann}]{calzetti00}
Calzetti, D., Armus, L., Bohlin, R.~C., {et~al.} 2000, ApJ, 533, 682

\bibitem[{Calzetti {et~al.}(1994)Calzetti, Kinney, \&
  Storchi-Bergmann}]{calzetti94}
Calzetti, D., Kinney, A.~L., \& Storchi-Bergmann, T. 1994, ApJ, 429, 582

\bibitem[{{Casey}(2012)}]{casey12}
{Casey}, C.~M. 2012, Monthly Notices of the Royal Astronomical Society, 425,
  3094

\bibitem[{Chary \& Elbaz(2001)}]{ce01}
Chary, R., \& Elbaz, D. 2001, ApJ, 556, 562

\bibitem[{{De Rossi} {et~al.}(2018){De Rossi}, {Rieke}, {Shivaei}, {Bromm}, \&
  {Lyu}}]{derossi18}
{De Rossi}, M.~E., {Rieke}, G.~H., {Shivaei}, I., {Bromm}, V., \& {Lyu}, J.
  2018, The Astrophysical Journal, 869, 4

\bibitem[{{Draine}(2011)}]{draine11}
{Draine}, B.~T. 2011, {Physics of the Interstellar and Intergalactic Medium}

\bibitem[{Draine {et~al.}(2007b)Draine, Dale, Bendo, Gordon, Smith, Armus,
  Engelbracht, Helou, Kennicutt, Li, Roussel, Walter, Calzetti, Moustakas,
  Murphy, Rieke, Bot, Hollenbach, Sheth, \& Teplitz}]{draine07b}
Draine, B.~T., Dale, D.~A., Bendo, G., {et~al.} 2007b, ApJ, 663, 866

\bibitem[{{Dunlop} {et~al.}(2017){Dunlop}, {McLure}, {Biggs}, {Geach},
  {Micha{\l}owski}, {Ivison}, {Rujopakarn}, {van Kampen}, {Kirkpatrick},
  {Pope}, {Scott}, {Swinbank}, {Targett}, {Aretxaga}, {Austermann}, {Best},
  {Bruce}, {Chapin}, {Charlot}, {Cirasuolo}, {Coppin}, {Ellis}, {Finkelstein},
  {Hayward}, {Hughes}, {Ibar}, {Jagannathan}, {Khochfar}, {Koprowski},
  {Narayanan}, {Nyland}, {Papovich}, {Peacock}, {Rieke}, {Robertson},
  {Vernstrom}, {Werf}, {Wilson}, \& {Yun}}]{dunlop17}
{Dunlop}, J.~S., {McLure}, R.~J., {Biggs}, A.~D., {et~al.} 2017, MNRAS, 466,
  861

\bibitem[{{Elbaz} {et~al.}(2011){Elbaz}, {Dickinson}, {Hwang},
  {D{\'{\i}}az-Santos}, {Magdis}, {Magnelli}, {Le Borgne}, {Galliano},
  {Pannella}, {Chanial}, {Armus}, {Charmandaris}, {Daddi}, {Aussel}, {Popesso},
  {Kartaltepe}, {Altieri}, {Valtchanov}, {Coia}, {Dannerbauer}, {Dasyra},
  {Leiton}, {Mazzarella}, {Alexander}, {Buat}, {Burgarella}, {Chary}, {Gilli},
  {Ivison}, {Juneau}, {Le Floc'h}, {Lutz}, {Morrison}, {Mullaney}, {Murphy},
  {Pope}, {Scott}, {Brodwin}, {Calzetti}, {Cesarsky}, {Charlot}, {Dole},
  {Eisenhardt}, {Ferguson}, {F{\"o}rster Schreiber}, {Frayer}, {Giavalisco},
  {Huynh}, {Koekemoer}, {Papovich}, {Reddy}, {Surace}, {Teplitz}, {Yun}, \&
  {Wilson}}]{elbaz11}
{Elbaz}, D., {Dickinson}, M., {Hwang}, H.~S., {et~al.} 2011, A\&A, 533, A119

\bibitem[{{Engelbracht} {et~al.}(2005){Engelbracht}, {Gordon}, {Rieke},
  {Werner}, {Dale}, \& {Latter}}]{engelbracht05}
{Engelbracht}, C.~W., {Gordon}, K.~D., {Rieke}, G.~H., {et~al.} 2005, ApJL,
  628, L29

\bibitem[{Erb {et~al.}(2006a)Erb, Shapley, Pettini, Steidel, Reddy, \&
  Adelberger}]{erb06a}
Erb, D.~K., Shapley, A.~E., Pettini, M., {et~al.} 2006a, ApJ, 644, 813

\bibitem[{{Franco} {et~al.}(2018){Franco}, {Elbaz}, {B{\'e}thermin},
  {Magnelli}, {Schreiber}, {Ciesla}, {Dickinson}, {Nagar}, {Silverman},
  {Daddi}, {Alexander}, {Wang}, {Pannella}, {Le Floc'h}, {Pope}, {Giavalisco},
  {Maury}, {Bournaud}, {Chary}, {Demarco}, {Ferguson}, {Finkelstein}, {Inami},
  {Iono}, {Juneau}, {Lagache}, {Leiton}, {Lin}, {Magdis}, {Messias},
  {Motohara}, {Mullaney}, {Okumura}, {Papovich}, {Pforr}, {Rujopakarn},
  {Sargent}, {Shu}, \& {Zhou}}]{franco18}
{Franco}, M., {Elbaz}, D., {B{\'e}thermin}, M., {et~al.} 2018, A\&A, 620, A152

\bibitem[{{Frayer} {et~al.}(2006){Frayer}, {Huynh}, {Chary}, {Dickinson},
  {Elbaz}, {Fadda}, {Surace}, {Teplitz}, {Yan}, \& {Mobasher}}]{frayer06}
{Frayer}, D.~T., {Huynh}, M.~T., {Chary}, R., {et~al.} 2006, The Astrophysical
  Journal, 647, L9

\bibitem[{{Goicoechea} {et~al.}(2016){Goicoechea}, {Pety}, {Cuadrado},
  {Cernicharo}, {Chapillon}, {Fuente}, {Gerin}, {Joblin}, {Marcelino}, \&
  {Pilleri}}]{goicoechea16}
{Goicoechea}, J.~R., {Pety}, J., {Cuadrado}, S., {et~al.} 2016, Nature, 537,
  207

\bibitem[{{Kashino} {et~al.}(2017){Kashino}, {Silverman}, {Sanders},
  {Kartaltepe}, {Daddi}, {Renzini}, {Valentino}, {Rodighiero}, {Juneau},
  {Kewley}, {Zahid}, {Arimoto}, {Nagao}, {Chu}, {Sugiyama}, {Civano}, {Ilbert},
  {Kajisawa}, {Le F{\`e}vre}, {Maier}, {Masters}, {Miyaji}, {Onodera},
  {Puglisi}, \& {Taniguchi}}]{kashino17}
{Kashino}, D., {Silverman}, J.~D., {Sanders}, D., {et~al.} 2017, ApJ, 835, 88

\bibitem[{{Kashino} {et~al.}(2019){Kashino}, {Silverman}, {Sanders},
  {Kartaltepe}, {Daddi}, {Renzini}, {Rodighiero}, {Puglisi}, {Valentino},
  {Juneau}, {Arimoto}, {Nagao}, {Ilbert}, {Le F{\`e}vre}, \&
  {Koekemoer}}]{kashino19}
---. 2019, The Astrophysical Journal Supplement Series, 241, 10

\bibitem[{Kennicutt \& Evans(2012)}]{kennicutt12}
Kennicutt, R.~C., \& Evans, N.~J. 2012, Annual Review of Astronomy and
  Astrophysics, 50, 531

\bibitem[{{Kewley} \& {Dopita}(2002)}]{kewley02}
{Kewley}, L.~J., \& {Dopita}, M.~A. 2002, ApJS, 142, 35

\bibitem[{{Kewley} {et~al.}(2015){Kewley}, {Zahid}, {Geller}, {Dopita},
  {Hwang}, \& {Fabricant}}]{kewley15}
{Kewley}, L.~J., {Zahid}, H.~J., {Geller}, M.~J., {et~al.} 2015, ApJL, 812, L20

\bibitem[{{Kimura} {et~al.}(2010){Kimura}, {Maihara}, {Iwamuro}, {Akiyama},
  {Tamura}, {Dalton}, {Takato}, {Tait}, {Ohta}, {Eto}, {Mochida}, {Elms},
  {Kawate}, {Kurakami}, {Moritani}, {Noumaru}, {Ohshima}, {Sumiyoshi}, {Yabe},
  {Brzeski}, {Farrell}, {Frost}, {Gillingham}, {Haynes}, {Moore}, {Muller},
  {Smedley}, {Smith}, {Bonfield}, {Brooks}, {Holmes}, {Curtis Lake}, {Lee},
  {Lewis}, {Froud}, {Tosh}, {Woodhouse}, {Blackburn}, {Content}, {Dipper},
  {Murray}, {Sharples}, \& {Robertson}}]{kiruma10}
{Kimura}, M., {Maihara}, T., {Iwamuro}, F., {et~al.} 2010, PASJ, 62, 1135

\bibitem[{Kriek {et~al.}(2015)Kriek, Shapley, Reddy, Siana, Coil, Mobasher,
  Freeman, de~Groot, Price, Sanders, Shivaei, Brammer, Momcheva, Skelton, van
  Dokkum, Whitaker, Aird, Azadi, Kassis, Bullock, Conroy, Dave, Keres, \&
  Krumholz}]{kriek15}
Kriek, M., Shapley, A.~E., Reddy, N.~A., {et~al.} 2015, ApJS, 218, 15

\bibitem[{{Kriek} {et~al.}(2019){Kriek}, {Price}, {Conroy}, {Suess}, {Mowla},
  {Pasha}, {Bezanson}, {van Dokkum}, \& {Barro}}]{kriek19}
{Kriek}, M., {Price}, S.~H., {Conroy}, C., {et~al.} 2019, The Astrophysical
  Journal, 880, L31

\bibitem[{{Kurt} \& {Dufour}(1998)}]{kurt98}
{Kurt}, C.~M., \& {Dufour}, R.~J. 1998, in Revista Mexicana de Astronomia y
  Astrofisica Conference Series, Vol.~7, Revista Mexicana de Astronomia y
  Astrofisica Conference Series, ed. R.~J. {Dufour} \& S.~{Torres-Peimbert},
  202

\bibitem[{{Liu} {et~al.}(2008){Liu}, {Shapley}, {Coil}, {Brinchmann}, \&
  {Ma}}]{liu08}
{Liu}, X., {Shapley}, A.~E., {Coil}, A.~L., {Brinchmann}, J., \& {Ma}, C.-P.
  2008, ApJ, 678, 758

\bibitem[{Madau \& Dickinson(2014)}]{madau14}
Madau, P., \& Dickinson, M. 2014, Annual Review of Astronomy and Astrophysics,
  52, 415

\bibitem[{{Magnelli} {et~al.}(2011){Magnelli}, {Elbaz}, {Chary}, {Dickinson},
  {Le Borgne}, {Frayer}, \& {Willmer}}]{magnelli11}
{Magnelli}, B., {Elbaz}, D., {Chary}, R.~R., {et~al.} 2011, A\&A, 528, A35

\bibitem[{{Marble} {et~al.}(2010){Marble}, {Engelbracht}, {van Zee}, {Dale},
  {Smith}, {Gordon}, {Wu}, {Lee}, {Kennicutt}, {Skillman}, {Johnson}, {Block},
  {Calzetti}, {Cohen}, {Lee}, \& {Schuster}}]{marble10}
{Marble}, A.~R., {Engelbracht}, C.~W., {van Zee}, L., {et~al.} 2010, The
  Astrophysical Journal, 715, 506

\bibitem[{{McLean} {et~al.}(2012){McLean}, {Steidel}, {Epps}, {Konidaris},
  {Matthews}, {Adkins}, {Aliado}, {Brims}, {Canfield}, {Cromer}, {Fucik},
  {Kulas}, {Mace}, {Magnone}, {Rodriguez}, {Rudie}, {Trainor}, {Wang}, {Weber},
  \& {Weiss}}]{mclean12}
{McLean}, I.~S., {Steidel}, C.~C., {Epps}, H.~W., {et~al.} 2012, in Society of
  Photo-Optical Instrumentation Engineers (SPIE) Conference Series, Vol. 8446,
  Society of Photo-Optical Instrumentation Engineers (SPIE) Conference Series,
  0

\bibitem[{{Nagao} {et~al.}(2011){Nagao}, {Maiolino}, {Marconi}, \&
  {Matsuhara}}]{nagao11}
{Nagao}, T., {Maiolino}, R., {Marconi}, A., \& {Matsuhara}, H. 2011, Astronomy
  and Astrophysics, 526, A149

\bibitem[{{Osterbrock} \& {Ferland}(2006)}]{osterbrock06}
{Osterbrock}, D.~E., \& {Ferland}, G.~J. 2006, {Astrophysics of gaseous nebulae
  and active galactic nuclei}

\bibitem[{{Pettini} \& {Pagel}(2004)}]{pp04}
{Pettini}, M., \& {Pagel}, B.~E.~J. 2004, MNRAS, 348, L59

\bibitem[{{Prevot} {et~al.}(1984){Prevot}, {Lequeux}, {Prevot}, {Maurice}, \&
  {Rocca-Volmerange}}]{prevot84}
{Prevot}, M.~L., {Lequeux}, J., {Prevot}, L., {Maurice}, E., \&
  {Rocca-Volmerange}, B. 1984, A\&A, 132, 389

\bibitem[{{Price} {et~al.}(2014){Price}, {Kriek}, {Brammer}, {Conroy},
  {F{\"o}rster Schreiber}, {Franx}, {Fumagalli}, {Lundgren}, {Momcheva},
  {Nelson}, {Skelton}, {van Dokkum}, {Whitaker}, \& {Wuyts}}]{price14}
{Price}, S.~H., {Kriek}, M., {Brammer}, G.~B., {et~al.} 2014, ApJ, 788, 86

\bibitem[{Reddy {et~al.}(2012a)Reddy, Dickinson, Elbaz, Morrison, Giavalisco,
  Ivison, Papovich, Scott, Buat, Burgarella, Charmandaris, Daddi, Magdis,
  Murphy, Altieri, Aussel, Dannerbauer, Dasyra, Hwang, Kartaltepe, Leiton,
  Magnelli, \& Popesso}]{reddy12a}
Reddy, N., Dickinson, M., Elbaz, D., {et~al.} 2012a, ApJ, 744, 154

\bibitem[{Reddy {et~al.}(2015)Reddy, Kriek, Shapley, Freeman, Siana, Coil,
  Mobasher, Price, Sanders, \& Shivaei}]{reddy15}
Reddy, N.~A., Kriek, M., Shapley, A.~E., {et~al.} 2015, ApJ, 806, 259

\bibitem[{{Rieke} {et~al.}(2009){Rieke}, {Alonso-Herrero}, {Weiner},
  {P{\'e}rez-Gonz{\'a}lez}, {Blaylock}, {Donley}, \& {Marcillac}}]{rieke09}
{Rieke}, G.~H., {Alonso-Herrero}, A., {Weiner}, B.~J., {et~al.} 2009, ApJ, 692,
  556

\bibitem[{{Sanders} {et~al.}(2016a){Sanders}, {Shapley}, {Kriek}, {Reddy},
  {Freeman}, {Coil}, {Siana}, {Mobasher}, {Shivaei}, {Price}, \& {de
  Groot}}]{sanders16a}
{Sanders}, R.~L., {Shapley}, A.~E., {Kriek}, M., {et~al.} 2016a, ApJ, 816, 23

\bibitem[{{Sanders} {et~al.}(2018){Sanders}, {Shapley}, {Kriek}, {Freeman},
  {Reddy}, {Siana}, {Coil}, {Mobasher}, {Dav{\'e}}, {Shivaei}, {Azadi},
  {Price}, {Leung}, {Fetherolf}, {de Groot}, {Zick}, {Fornasini}, \&
  {Barro}}]{sanders18}
---. 2018, ApJ, 858, 99

\bibitem[{{Sanders} {et~al.}(2019){Sanders}, {Shapley}, {Reddy}, {Kriek},
  {Siana}, {Coil}, {Mobasher}, {Shivaei}, {Freeman}, {Azadi}, {Price}, {Leung},
  {Fetherolf}, {de Groot}, {Zick}, {Fornasini}, \& {Barro}}]{sanders19}
{Sanders}, R.~L., {Shapley}, A.~E., {Reddy}, N.~A., {et~al.} 2019, arXiv
  e-prints, arXiv:1907.00013

\bibitem[{{Savage} \& {Mathis}(1979)}]{savage79}
{Savage}, B.~D., \& {Mathis}, J.~S. 1979, ARA\&A, 17, 73

\bibitem[{{Schreiber} {et~al.}(2018){Schreiber}, {Elbaz}, {Pannella}, {Ciesla},
  {Wang}, \& {Franco}}]{schreiber18}
{Schreiber}, C., {Elbaz}, D., {Pannella}, M., {et~al.} 2018, Astronomy and
  Astrophysics, 609, A30

\bibitem[{{Sharples} {et~al.}(2013){Sharples}, {Bender}, {Agudo Berbel},
  {Bezawada}, {Castillo}, {Cirasuolo}, {Davidson}, {Davies}, {Dubbeldam},
  {Fairley}, {Finger}, {F{\"o}rster Schreiber}, {Gonte}, {Hess}, {Jung},
  {Lewis}, {Lizon}, {Muschielok}, {Pasquini}, {Pirard}, {Popovic}, {Ramsay},
  {Rees}, {Richter}, {Riquelme}, {Rodrigues}, {Saviane}, {Schlichter},
  {Schmidtobreick}, {Segovia}, {Smette}, {Szeifert}, {van Kesteren}, {Wegner},
  \& {Wiezorrek}}]{sharples13}
{Sharples}, R., {Bender}, R., {Agudo Berbel}, A., {et~al.} 2013, The Messenger,
  151, 21

\bibitem[{{Shipley} {et~al.}(2016){Shipley}, {Papovich}, {Rieke}, {Brown}, \&
  {Moustakas}}]{shipley16}
{Shipley}, H.~V., {Papovich}, C., {Rieke}, G.~H., {Brown}, M.~J.~I., \&
  {Moustakas}, J. 2016, ApJ, 818, 60

\bibitem[{Shivaei {et~al.}(2015a)Shivaei, Reddy, Steidel, \&
  Shapley}]{shivaei15a}
Shivaei, I., Reddy, N.~A., Steidel, C.~C., \& Shapley, A.~E. 2015a, ApJ, 804,
  149

\bibitem[{Shivaei {et~al.}(2015b)Shivaei, Reddy, Shapley, Kriek, Siana,
  Mobasher, Coil, Freeman, Sanders, Price, de~Groot, \& Azadi}]{shivaei15b}
Shivaei, I., Reddy, N.~A., Shapley, A.~E., {et~al.} 2015b, ApJ, 815, 98

\bibitem[{{Shivaei} {et~al.}(2016){Shivaei}, {Kriek}, {Reddy}, {Shapley},
  {Barro}, {Conroy}, {Coil}, {Freeman}, {Mobasher}, {Siana}, {Sanders},
  {Price}, {Azadi}, {Pasha}, \& {Inami}}]{shivaei16}
{Shivaei}, I., {Kriek}, M., {Reddy}, N.~A., {et~al.} 2016, ApJL, 820, L23

\bibitem[{{Shivaei} {et~al.}(2017){Shivaei}, {Reddy}, {Shapley}, {Siana},
  {Kriek}, {Mobasher}, {Coil}, {Freeman}, {Sanders}, {Price}, {Azadi}, \&
  {Zick}}]{shivaei17}
{Shivaei}, I., {Reddy}, N.~A., {Shapley}, A.~E., {et~al.} 2017, ApJ, 837, 157

\bibitem[{{Shivaei} {et~al.}(2018){Shivaei}, {Reddy}, {Siana}, {Shapley},
  {Kriek}, {Mobasher}, {Freeman}, {Sanders}, {Coil}, {Price}, {Fetherolf},
  {Azadi}, {Leung}, \& {Zick}}]{shivaei18}
{Shivaei}, I., {Reddy}, N.~A., {Siana}, B., {et~al.} 2018, The Astrophysical
  Journal, 855, 42

\bibitem[{{Smith} {et~al.}(2019){Smith}, {Armus}, {Dav{\'e}}, {Ferkinhoff},
  {Groves}, {Kewley}, {Murphy}, {Pope}, {Shivaei}, \& {Skillman}}]{smith19}
{Smith}, J.~D., {Armus}, L., {Dav{\'e}}, R., {et~al.} 2019, Bulletin of the
  American Astronomical Society, 51, 400

\bibitem[{{Steidel} {et~al.}(2016){Steidel}, {Strom}, {Pettini}, {Rudie},
  {Reddy}, \& {Trainor}}]{steidel16}
{Steidel}, C.~C., {Strom}, A.~L., {Pettini}, M., {et~al.} 2016, ApJ, 826, 159

\bibitem[{{Strom} {et~al.}(2017{\natexlab{a}}){Strom}, {Steidel}, {Rudie},
  {Trainor}, {Pettini}, \& {Reddy}}]{strom18}
{Strom}, A.~L., {Steidel}, C.~C., {Rudie}, G.~C., {et~al.} 2017{\natexlab{a}},
  The Astrophysical Journal, 836, 164

\bibitem[{{Strom} {et~al.}(2017{\natexlab{b}}){Strom}, {Steidel}, {Rudie},
  {Trainor}, {Pettini}, \& {Reddy}}]{strom17}
---. 2017{\natexlab{b}}, ApJ, 836, 164

\bibitem[{{Theios} {et~al.}(2019){Theios}, {Steidel}, {Strom}, {Rudie},
  {Trainor}, \& {Reddy}}]{theios19}
{Theios}, R.~L., {Steidel}, C.~C., {Strom}, A.~L., {et~al.} 2019, The
  Astrophysical Journal, 871, 128

\bibitem[{{Whitaker} {et~al.}(2017){Whitaker}, {Pope}, {Cybulski}, {Casey},
  {Popping}, \& {Yun}}]{whitaker17}
{Whitaker}, K.~E., {Pope}, A., {Cybulski}, R., {et~al.} 2017, ArXiv e-prints
  1710.06872, arXiv:1710.06872

\bibitem[{Whitaker {et~al.}(2014b)Whitaker, Franx, Leja, van Dokkum, Henry,
  Skelton, Fumagalli, Momcheva, Brammer, Labb\'{e}, Nelson, \&
  Rigby}]{whitaker14b}
Whitaker, K.~E., Franx, M., Leja, J., {et~al.} 2014b, ApJ, 795, 104

\bibitem[{{Wisnioski} {et~al.}(2015){Wisnioski}, {F{\"o}rster Schreiber},
  {Wuyts}, {Wuyts}, {Bandara}, {Wilman}, {Genzel}, {Bender}, {Davies},
  {Fossati}, {Lang}, {Mendel}, {Beifiori}, {Brammer}, {Chan}, {Fabricius},
  {Fudamoto}, {Kulkarni}, {Kurk}, {Lutz}, {Nelson}, {Momcheva}, {Rosario},
  {Saglia}, {Seitz}, {Tacconi}, \& {van Dokkum}}]{wisnioski15}
{Wisnioski}, E., {F{\"o}rster Schreiber}, N.~M., {Wuyts}, S., {et~al.} 2015,
  The Astrophysical Journal, 799, 209

\bibitem[{Wuyts {et~al.}(2008)Wuyts, Labb\'{e}, Schreiber, Franx, Rudnick,
  Brammer, \& van Dokkum}]{wuyts08}
Wuyts, S., Labb\'{e}, I., Schreiber, N. M.~F., {et~al.} 2008, ApJ, 682, 985

\end{thebibliography}

\end{document}